\documentclass[a4paper]{jps-cp}
\usepackage[dvipdfmx]{graphicx}
\usepackage{bm,amssymb}
\usepackage{amsmath}
\usepackage{txfonts}
\usepackage{braket}
\usepackage[table]{xcolor}
\usepackage{arydshln}

\title{RKKY interactions of CeB$_6$ based on effective Wannier model}

\author{Takemi Yamada and Katsurou Hanzawa}

\inst{Department of Physics, Faculty of Science and Technology, Tokyo University of Science, Chiba, Noda, 278-8510, Japan}

\email{t-yamada@rs.tus.ac.jp}

\recdate{September 17, 2019}

\abst{
We examine the RKKY interactions of CeB$_6$ between multipole moments based on the effective Wannier model obtained from the bandstructure calculation including 14 Ce-$f$ orbitals and 60 conduction orbitals of Ce-$d,s$ and B-$p,s$. 
By using the $f$-$c$ mixing matrix elements of the Wannier model together with the conduction band dispersion, 
the multipole couplings with the RKKY oscillation are obtained for the active moments in $\Gamma_{8}$ subspace. %in the wavevector $\bm{q}$-space. 
Both of the $\Gamma_{5g}$ quadrupole $O_{xy}$ and the $\Gamma_{2u}$ octupole $T_{xyz}$ couplings are largely enhanced with $\bm{q}=(\pi,\pi,\pi)$ which naturally explains the antiferro-quadrupolar phase of the phase II, and are also enhanced with $\bm{q}=(0,0,0)$ corresponding to the elastic softening of $C_{44}$. 
Also the couplings of the $\Gamma_{5u}$ octupole $T_{z}^{\beta}$ %and the $\Gamma_{3g}$ quadrupole $O_{x^2-y^2}$ 
is quite large for $\bm{q}=(0,0,\pi)$ which is related to the antiferro-octupolar ordering of a possible candidate for the phase IV of Ce$_{x}$La$_{1-x}$B$_6$. 
}

\kword{CeB$_6$, RKKY interaction, multipole, bandstructure calculation}

\begin{document}
\maketitle

\section{Introduction}
Electronic state of CeB$_6$ has been one of the central issues in the heavy Fermion systems, 
since it exhibits a rich phase diagram of the multipole orderings\cite{Cameron2016,Thalmeier2019} 
due to the $\Gamma_8$ quartet ground state with the degrees of freedom of multipole moments as shown in Table \ref{table1}. 
Several multipole orderings have been observed in temperature and magnetic field $(T,H)$ phase diagram such as the antifero-quadrupolar (AFQ) ordering of $\Gamma_{5g}$ quadrupole moments $(O_{yz},O_{zx},O_{xy})$ with a critical transition temperature $T_{Q}=3.2$ K (phase II) and the antifero-magnetic ordering of $\Gamma_{4u}$ magnetic multipoles $(\sigma^{x},\sigma^{y},\sigma^{z})$ with $T_{N}=2.3$ K (phase III). The antifero-octupolar (AFO) ordering of $\Gamma_{5u}$ octupoles $(T^{\beta}_{x},T^{\beta}_{y},T^{\beta}_{z})$ is also discussed as a possible candidate of the phase IV in the La-substitution system of Ce$_x$La$_{1-x}$B$_6$. 

The 4$f$ electrons of CeB$_6$ are almost localized from several experiments. 
The Fermi-surface (FS) has been observed in the de Haas-van Alphen (dHvA) experiments\cite{Onuki1989}, 
the angle resolved photoemission spectroscopy (ARPES)\cite{Neupane2015,Ramankuttya2016} 
and the high-resolution photoemission tomography\cite{Koitzsch2016}, 
where an ellipsoidal FS centered at X point in the Brillouin zone (BZ) has been confirmed
and is almost the same as that of LaB$_6$ with the 4$f^{0}$ state. 
In such a localized $f$ electron system, Ruderman-Kittel-Kasuya-Yosida (RKKY) interaction\cite{RK1954,Kasuya1956,Yosida1957} 
plays an important role for the multipole ordering and must determine the ordering moments and wavevectors. 
The phonomenological RKKY models of CeB$_6$\cite{Ohkawa1983,Shiina1997} 
succeeded in reproducing the basic phase diagram in $T$-$H$ plane 
and giving a great advance in the multipole physics in $\Gamma_{8}$ ground state system. 
%and gave a great insight of the multipole physics in $\Gamma_{8}$ ground state system. 

However in these studies, only the nearest neighbor RKKY Hamiltonian with a symmetric coupling 
and/or asymmetric correction terms was used for explaining the experimental phase diagram. 
As shown in the original studies\cite{RK1954,Kasuya1956,Yosida1957}, 
the RKKY interaction has a decaying and oscillating function of $2k_{\rm F}R$ 
with the Fermi wavenumber $k_{\rm F}$ of conduction $(c)$ band and distance between multipole moments $R$, 
and thus these effects should be taken into account through 
the microscopic description of the $f$-$c$ mixing and $c$ band states.
%was investigated in detail, where all 15 active multipole moments had been taken into account properly, 
%within only nearest neighbor couplings is assumed and the couplings of $\Gamma_{5g}$ quadrupoles are considered to be largest among all multipole moments. 
%where only nearest neighbor couplings and the largest $\Gamma_{5g}$ quadrupoles couplings ware assumed. 
%This assumption was discussed from the symmetry of the RKKY couplings\cite{Shiba1999,Hanzawa2000}, 
%but there was no explicit calculation for the signs and values of the couplings, 
The explicit derivation of the RKKY multipole couplings based on the realistic bandstructure calculation
%and its microscopic understanding of the multipole order 
still has been an important challenge for the microscopic understanding of the multipole order. 
%Recently, we have studied the electronic states of CeB$_6$ and RKKY interactions based on the realistic bandstructure calculation together with the microscopic Wannier orbital models

%As is often discussed in the RKKY mechanism, the $c$ band states and their couplings with the $f$ states in the realistic materials must be important for determining the ordering moment types and wavevectors. 
%Therefore the microscopic description of the $c$ band states and $f$-$c$ mixing elements from the realistic bandstructure calculation is needed, though such studies are quite limited\cite{Tanaka2010,Hanzawa2015}. 
%In these studies\cite{Tanaka2010,Hanzawa2015}, the $c$ states is described by the Wannier orbitals obtained from the bandstructure calculation but the $f$-$c$ mixing elements using the calculation of the RKKY coupling are treated by the SK parameters only with the nearest neighbor sites, where several arbitrary parameters and assumptions are included. 
%Hence more decisive and widely-applicable approach reflecting the individual material properties is highly desired. 

In this study, we derive the RKKY interaction of CeB$_6$ based on the 74-orbital  effective Wannier model derived from the bandstructure calculation directly\cite{YH2019}. By using the realistic $c$ band dispersion together with the $f$-$c$ mixing matrix elements from the Wannier model of CeB$_6$, we calculate the RKKY couplings between the active multipole moments in $\Gamma_8$ subspace mediated by the realistic $c$ band states explicitly.
The obtained RKKY multipole interactions show that the 1st leading multipole mode is the $\bm{q}=(\pi,\pi,\pi)$-AFQ ordering with $\Gamma_{5g}$ quadrupole $O_{xy}$ together with $\Gamma_{2u}$ octupole $T_{xyz}$ 
and the 2nd leading mode is the $\bm{q}=(0,0,\pi)$-AFO ordering with $\Gamma_{5u}$ octupole $T_{z}^{\beta}$. 
%The former (latter) corresponds to the instability towards the phase II (phase IV) 
%and main results obtained here can naturally explain the observed phase II (AFQ) and the phase IV (AFO). 

%Table.1%%%%%%%%%%%%%%%%%%%%%%%%%%%%%%%%%%%%%%%%%%%%%%%%%%%%%%%%%%%%%%%%%%%%%%%%%%%%%%%%%%%%%%%%
\begin{table}[t]
%\vspace{-0.4cm}
\centering
%\scalebox{0.89}{
\begin{tabular}{cccc}
\hline \rule{0pt}{4mm}
IRR & notation & pseudospin rep. & multipole \\[2pt] \hline\hline \rule{0pt}{4mm}
$\Gamma_{2u}$ & $\xi$ & $\tau^{y}$ & $\frac{2}{9\sqrt{5}}T_{xyz}$ \\[2pt] \rule{0pt}{4mm}
$\Gamma_{3g}$ & $\bm{\tau}\rq{}=(\tau^z,\tau^x)$ & $(\tau^{z},\tau^{x})$ & $\frac{1}{4}(O_{u},O_{v})$ \\[2pt] \rule{0pt}{4mm}
$\Gamma_{5g}$ & $\bm{\mu}=(\mu^x,\mu^y,\mu^z)$ & $(\tau^{y}\sigma^{x},\tau^{y}\sigma^{y},\tau^{y}\sigma^{z})$ & $(O_{yz},O_{zx},O_{xy})$ \\[2pt] \rule{0pt}{4mm}
$\Gamma_{4u}^{(1)}$ & $\bm{\sigma}=(\sigma^x,\sigma^y,\sigma^z)$ &$(\sigma^{x},\sigma^{y},\sigma^{z})$ &$\frac{14}{15}\bm{J}-\frac{4}{45}\bm{T}^{\alpha}$ \\[2pt] \rule{0pt}{4mm}
$\Gamma_{4u}^{(2)}$ & $\bm{\eta}=(\eta^x,\eta^y,\eta^z)$ &$(\eta^{+}\sigma^{x},\eta^{-}\sigma^{y},\tau^{z}\sigma^{z})$ &$-\frac{2}{15}\bm{J}+\frac{7}{45}\bm{T}^{\alpha}$ \\[2pt] \rule{0pt}{4mm}
$\Gamma_{5u}$  & $\bm{\zeta}=(\zeta^x,\zeta^y,\zeta^z)$& $(\zeta^{+}\sigma^{x},\zeta^{-}\sigma^{y},\tau^{x}\sigma^{z})$ & $\frac{1}{3\sqrt{5}}\bm{T}^{\beta}$ \\[2pt]
\hline
\end{tabular}
%}
\vspace{0.25cm}
\caption{
The irreducible representations (IRRs), notations and pseudospin representations for the active multipole moments in $\Gamma_8$ subspace
%\cite{Kusunose2008} %,Shiina1997,Shiina1998}
where 
$\bm{J}~(\bm{T}^{\alpha,\beta})$ is the dipole (octupole), 
$\bm{J}=(J_{x},J_{y},J_{z})$, 
$\bm{T}^{\alpha(\beta)}=(T_{x}^{\alpha(\beta)},T_{y}^{\alpha(\beta)},T_{z}^{\alpha(\beta)})$, 
$\eta^{\pm}=-\frac{1}{2}(\tau^{z}\mp\sqrt{3}\tau^{x})$, 
$\zeta^{\pm}=-\frac{1}{2}(\tau^{x}\pm\sqrt{3}\tau^{z})$, 
and $g~(u)$ means even (odd) time-reversal symmetry. 
Here all the multipole operators are normalized. 
}\label{table1}
\end{table} 
%%%%%%%%%%%%%%%%%%%%%%%%%%%%%%%%%%%%%%%%%%%%%%%%%%%%%%%%%%%%%%%%%%%%%%%%%%%%%%%%%%%%%%%%%%%%%%%

%=====| Model & Formulation |==================================================================
\section{Model \& Formulation}
%\subsection{Bandstructure calculation \& Wannier model}
First we perform the bandstructure calculation of CeB$_6$ and LaB$_6$ 
by using the WIEN2k code\cite{w2k2002}, 
which is based on the density-functional theory (DFT) and includes the effect of the spin-orbit coupling (SOC) within the second variation approximation. 
The explicit bandstructures and FSs are shown in Fig.1 (a)-(d) of Ref. \cite{YH2019}, 
where three FSs are obtained in CeB$_6$, %which is inconsistent with the experiments
while an ellipsoidal FS centered at X point %slightly connected each other 
is obtained in LaB$_6$, 
%which is consistent with the experimental results\cite{Onuki1989,Neupane2015,Ramankuttya2016,Koitzsch2016}.
which well accounts for the experimental results\cite{Onuki1989,Neupane2015,Ramankuttya2016,Koitzsch2016}.

%Except for the $f$ band states, the global bandstructures of CeB$_6$ and LaB$_6$ are closely resembled below and above $E_{\rm F}$. 
%The calculated FSs of CeB$_6$ and LaB$_6$ are plotted in Fig.\ref{Fig1} (b) and (d), respectively. 
%Three FSs are obtained from the 21th, 23th and 25th bands for CeB$_6$ 
%while for LaB$_6$ an ellipsoidal FS centered at X point slightly connected each other is obtained from the 21th-band. 
%Here we note that all bands have two-folded degeneracy due to the time-reversal symmetry 
%and two additional bands (1st and 2nd bands) are located in $E_{\rm F}-15$ eV (not shown) which are the lowest bands in the Wannier model. 

Next we construct the 74-orbital effective Wannier model based on the maximally localized Wannier functions (MLWFs) method\cite{w90-Mostofi2008,w2w-Kunes2010} from the DFT bandstructure of CeB$_6$, 
where 14 $f$-states from Ce-$f$ (7 orbital $\times$ 2 spin) 
and 60 $c$-states from Ce-$d$ (5 orbital $\times$ 2 spin), Ce-$s$ (1 orbital $\times$ 2 spin), B-$p$ (6 site $\times$ 3 orbital $\times$ 2 spin) and B-$s$ (6 site $\times$ 1 orbital $\times$ 2 spin) are fully included and the obtained bandstructure well reproduces the DFT band as shown in Fig.1 (e) and (f) of Ref. \cite{YH2019}. 
The obtained tight-binding (TB) Hamiltonian is given by the following form as, 
%\begin{align}
%H_{\rm TB}&=\sum_{ij}\sum_{mm\rq{}}h_{im,jm\rq{}}^{ff}f_{im}^{\dagger}f_{jm\rq{}}^{} +\sum_{ij}\sum_{\ell\ell\rq{}}h_{i\ell,j\ell\rq{}}^{cc}c_{i\ell}^{\dagger}c_{j\ell\rq{}}^{}%\nonumber\\
%+\sum_{ij}\sum_{m\ell}\left(V_{im,j\ell}f_{im}^{\dagger}c_{j\ell}^{}+h.c.\right),
%\label{eq:HTB}
%\end{align}
%where $f_{im}^{\dagger}~(c_{i\ell}^{\dagger})$ is a creation operator for a $f~(c)$ electron with unit-cell $i$
%and 14 (60) spin-orbital states $m~(\ell)$. 
%Here 14 $f$ states of $m$ are represented by the CEF eigenstates as $\Gamma_{8}$ quartet and $\Gamma_{7}$ doublet with the total angular momentum $J=5/2$, and $\Gamma_{6}$, $\Gamma_{7}$ doublets and $\Gamma_{8}$ quartet with $J=7/2$. 
%The $f$-$f$ ($c$-$c$) matrix element of $h_{im,jm\rq{}}^{ff}~(h_{i\ell,j\ell\rq{}}^{cc})$ 
%includes the $f~(c)$ energy levels, SOC couplings, CEF splittings and $f$-$f$ ($c$-$c$) hopping integrals, 
%and $V_{im,j\ell}$ is the $f$-$c$ mixing element which is finite only for the intersite terms due to the inversion symmetry. 
%The wavevector $\bm{k}$-representation of $H_{\rm TB}$ is given by, 
\begin{align}
H_{\rm TB}&=\sum_{\bm{k}}\sum_{mm\rq{}}h_{mm\rq{}}^{ff}(\bm{k})f_{\bm{k}m}^{\dagger}f_{\bm{k}m\rq{}}^{} +\sum_{\bm{k}}\sum_{\ell\ell\rq{}}h_{\ell\ell\rq{}}^{cc}(\bm{k})c_{\bm{k}\ell}^{\dagger}c_{\bm{k}\ell\rq{}}^{}%\nonumber\\
+\sum_{\bm{k}}\sum_{m\ell}\left(V_{\bm{k}m\ell}f_{\bm{k}m}^{\dagger}c_{\bm{k}\ell}^{}+h.c.\right)
%=\sum_{\bm{k}s}\varepsilon_{\bm{k}s}a_{\bm{k}s}^{\dagger}a_{\bm{k}s}^{},
\label{eq:HTB-k}
\end{align}
where $f_{\bm{k}m}^{\dagger}~(c_{\bm{k}\ell}^{\dagger})$ is a creation operator for a $f~(c)$ electron with wavevector $\bm{k}$
and 14 (60) spin-orbital states $m~(\ell)$. 
Here 14 $f$ states of $m$ are represented by the CEF eigenstates as $\Gamma_{8}$ quartet and $\Gamma_{7}$ doublet with the total angular momentum $J=5/2$, and $\Gamma_{6}$, $\Gamma_{7}$ doublets and $\Gamma_{8}$ quartet with $J=7/2$. 
The $f$-$f$ [$c$-$c$] matrix element of $h_{mm\rq{}}^{ff}(\bm{k})~[h_{\ell\ell\rq{}}^{cc}(\bm{k})]$ 
includes the $f~(c)$ energy levels, SOC couplings, CEF splittings and $f$-$f$ ($c$-$c$) hopping integrals, 
and $V_{\bm{k}m\ell}$ is the $f$-$c$ mixing element. %which is finite only for the intersite terms due to the inversion symmetry. 

\begin{figure*}[t]
\centering
\includegraphics[width=15.5cm]{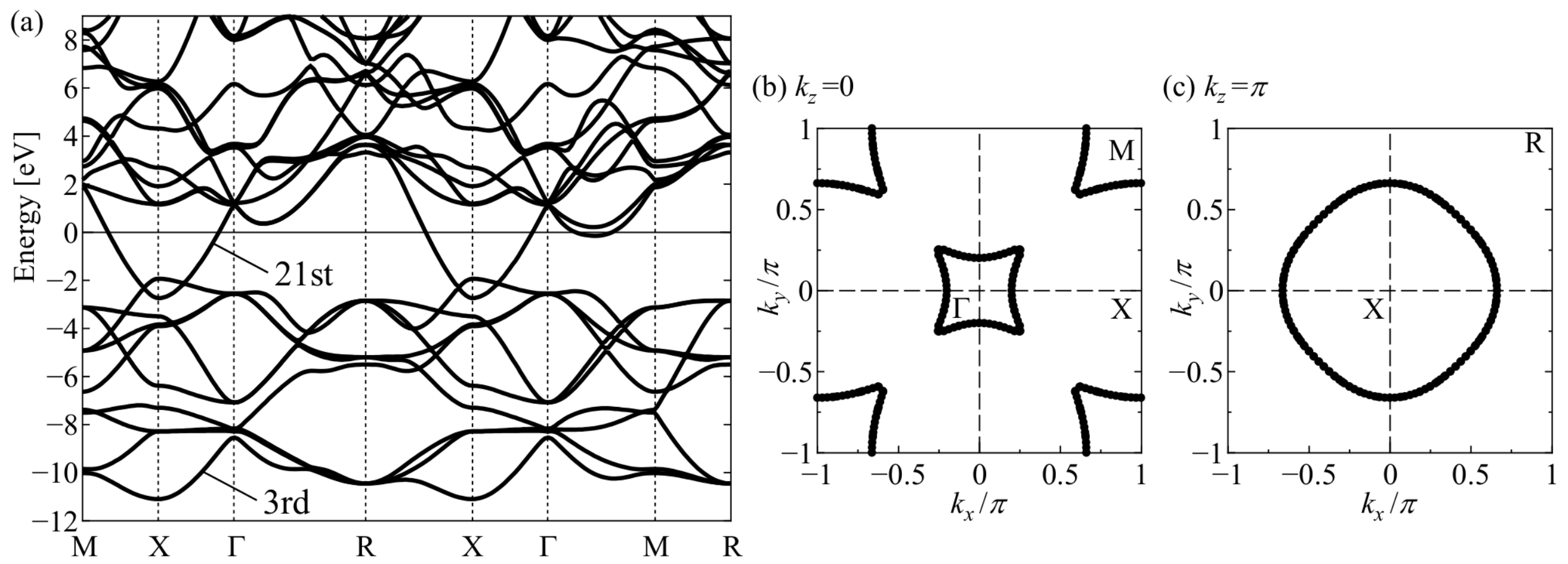}
%\vspace{-0.3cm}
\caption{%(Color online) 
(a) Bandstructure and (b),(c) FSs (21st band) of the conduction states in the Wannier model for CeB$_6$ in the simple cubic BZ, 
where the high-symmetry points are $\Gamma$[$(0,0,0)$], X[$(\pi,0,0)$], M[$(\pi.\pi,0)$] and R[$(\pi.\pi,\pi)$]. 
All bands have two-folded degeneracy due to the time-reversal symmetry and 1st and 2nd bands are located in $-15$ eV (not shown). 
}
\label{Fig1}
%\vspace{+0.2cm}
\end{figure*}

%\subsection{Derivation of RKKY Hamiltonian}
Here we consider the RKKY interaction between the multipole moments of $\Gamma_8$ quartet. 
For this purpose, we start from the localized $f$ limit where 
a $4f^{1}$ state is realized in $\Gamma_{8}$ at each Ce site and the $f$-$f$ hopping and $f$-$c$ mixing become zero. 
Hence %the remained bandstructure is consist of $c$ electron which is 
the remained $c$ electron Hamiltonian $H_{\rm TB}^{c}$ is diagonalized as follows, 
\begin{align}
H_{\rm TB}^{c}
=\sum_{\bm{k}}\sum_{\ell\ell\rq{}}h_{\ell\ell\rq{}}^{cc}(\bm{k})c_{\bm{k}\ell}^{\dagger}c_{\bm{k}\ell\rq{}}^{}
=\sum_{\bm{k}s}\varepsilon_{\bm{k}s}^{c}a_{\bm{k}s}^{\dagger}a_{\bm{k}s}^{},
\label{eq:HTBc-k}
\end{align}
where $a_{\bm{k}s}^{\dagger}$ is a creation operator for a electron with $\bm{k}$ and band-index $s$ 
with the $c$ band dispersion $\varepsilon_{\bm{k}s}^{c}$ and the eigenvector $u_{\bm{k}s\ell}^{c}$ where $%\displaystyle
a_{\bm{k}s}=\sum_{\ell}u_{\bm{k}s\ell}^{c}c_{\bm{k}\ell}$. 
%The obtained $\varepsilon_{\bm{k}s}^{c}$ and FSs are shown in 
Figure \ref{Fig1} shows the $c$ bandstructure [Fig. \ref{Fig1}(a)] and their FSs of the 21st band [Fig. \ref{Fig1}(b) \& (c)] 
and the obtained $c$ state well describes the experimental FSs\cite{Onuki1989,Neupane2015,Ramankuttya2016,Koitzsch2016} and 
is also almost same as the DFT-bandstructure of LaB$_6$ without $f$ electron. 

%Fig.1%%%%%%%%%%%%%%%%%%%%%%%%%%%%%%%%%%%%%%%%%%%%%%%%%%%%%%%%%%%%%%%%%%%%%%%%%%%%%%%%%%%%%%%%
%%%%%%%%%%%%%%%%%%%%%%%%%%%%%%%%%%%%%%%%%%%%%%%%%%%%%%%%%%%%%%%%%%%%%%%%%%%%%%%%%%%%%%%%%%%%%%%

%but use the $f$-$c$ mixing of the original Wannier model, 
By using the second-order perturbation w. r. t. the $f$-$c$ mixing $V_{\bm{k}m\ell}$ in the third term of $H_{\rm TB}$, 
we obtain the multi-orbital Kondo lattice Hamiltonian which is given by, 
\begin{align}
H_{\rm MKL}&=
%\sum_{im}\varepsilon_{\Gamma_8}^{f}f_{im}^{\dagger}f_{im}^{}
\sum_{im}
\left(\varepsilon_{\Gamma_{8}}^{f}+\Delta\varepsilon_{\Gamma_{8}}^{f}\right)
%\tilde{\varepsilon}_{\Gamma_{8}}^{f}
f_{im}^{\dagger}f_{im}^{}
+\sum_{\bm{k}s}\varepsilon_{\bm{k}s}^{c}a_{\bm{k}s}^{\dagger}a_{\bm{k}s}^{}
%+\sum_{\bm{k}\ell\ell\rq{}}h_{\ell\ell\rq{}}^{cc}(\bm{k})c_{\bm{k}\ell}^{\dagger}c_{\bm{k}\ell\rq{}}^{}%\nonumber\\
+\sum_{i}\sum_{mm\rq{}}\sum_{\bm{k}\bm{k\rq{}}}\sum_{\ell\ell\rq{}}J_{imm\rq{}}^{\bm{k}\ell,
\bm{k}\rq{}\ell\rq{}}f_{im}^{\dagger}f_{im\rq{}}^{}c_{\bm{k}\ell}^{\dagger}c_{\bm{k}\rq{}\ell\rq{}}^{},
\end{align}
where $f_{im}^{\dagger}$ is a creation operator for a $f$ electron with a Ce-atom $i$
and
%$m$ represents 
4-states in $\Gamma_8$ quartet $\Ket{m}=\Ket{1}\sim \Ket{4}$ 
%with an degenerate level $\tilde{\varepsilon}_{m}^{f}$, 
which are given with the $J_{z}$-base of $J=5/2$ $\Ket{M}$ explicitly as,
%$\Ket{1}=-\sqrt{\frac{1}{6}}\Ket{+\frac{3}{2}}-\sqrt{\frac{5}{6}}\Ket{-\frac{5}{2}}$, 
%$\Ket{2}=\Ket{+\frac{1}{2}}$, 
%$\Ket{3}=-\Ket{-\frac{1}{2}}$ and 
%$\Ket{4}=\sqrt{\frac{1}{6}}\Ket{-\frac{3}{2}}+\sqrt{\frac{5}{6}}\Ket{+\frac{5}{2}}$. 
\begin{align}
&\Ket{1}=-\sqrt{\frac{1}{6}}\Ket{+\frac{3}{2}}-\sqrt{\frac{5}{6}}\Ket{-\frac{5}{2}}\\
&\Ket{2}=+\Ket{+\frac{1}{2}}\\
&\Ket{3}=-\Ket{-\frac{1}{2}}\\
&\Ket{4}=+\sqrt{\frac{1}{6}}\Ket{-\frac{3}{2}}+\sqrt{\frac{5}{6}}\Ket{+\frac{5}{2}}, 
\end{align}
where $\varepsilon_{\Gamma_8}^{f}$ is the bare $f$ energy-level and 
%$\Delta\varepsilon_{\Gamma_{8}}^{f}$ is a energy shift so as to satisfy $n^{f}=1$ at each site 
$\Delta\varepsilon_{\Gamma_{8}}^{f}$ is a energy shift due to the DFT potential which is of the order of a few eV. 
%The $c$-$c$ matrix element $h_{\ell\ell\rq{}}^{cc}(\bm{k})$ 
%includes the $c$ orbital energy $\varepsilon_{\bm{k}\ell}^{c}$ for $\ell=\ell\rq{}$ and the $c$-$c$ hopping $t_{\ell\ell\rq{}}^{cc}(\bm{k})$ for $\ell\neq\ell\rq{}$. 
%The second term is rewritten by the $c$ band eigenstate 
%$\displaystyle c_{\bm{k}s}=\sum_{\ell}u_{\bm{k}s\ell}^{c}c_{\bm{k}\ell}$ with the eigenenergy $\varepsilon_{\bm{k}s}$ and eigenvector $u_{\bm{k}s\ell}^{c}$. 
The Kondo coupling $J_{imm\rq{}}^{\bm{k}\ell,\bm{k}\rq{}\ell\rq{}}$ can be written by the following simple form, 
\begin{align}
&J_{imm\rq{}}^{\bm{k}\ell,\bm{k}\rq{}\ell\rq{}}
=\frac{2}{N}
\frac{V_{\bm{k}m\ell}V_{\bm{k}\rq{}m\rq{}\ell\rq{}}^{*}}{\mu-\varepsilon_{\Gamma_8}^{f}}
e^{-i(\bm{k}-\bm{k}\rq{})\cdot\bm{R}_i}, 
\end{align}
where only $f^0$-intermediate process is considered 
%and the $f^2$-process contribution is same as that of $f^0$-process 
and the scattered $c$ orbital energies are fixed to $\mu$. 

The RKKY Hamiltonian can be obtained from the second-order perturbation w. r. t. the third term of $H_{\rm MKL}$ together with the thermal average for the $c$ states. The final form is given by, 
\begin{align}
&H_{\rm RKKY}\!=\!-\!\sum_{\Braket{ij}}\!\sum_{m_1m_2}\!\sum_{m_3m_4}\!
K_{m_1m_2m_3m_4}(\bm{R}_{ij})
f_{im_1}^{\dagger}
f_{im_2}^{}
f_{jm_4}^{\dagger}
f_{jm_3}^{}
\label{eq:HRKKY-r},\\
&K_{m_1m_2m_3m_4}(\bm{R}_{ij})=\frac{1}{N}\sum_{\bm{q}}K_{m_1m_2m_3m_4}(\bm{q})~e^{i\bm{q}\cdot(\bm{R}_i-\bm{R}_j)},\label{eq:Kij}
\end{align}
where $K_{m_1m_2m_3m_4}(\bm{R}_{ij})$ is the RKKY coupling between $\{m_1,m_2\}$ at Ce-atom $\bm{R}_i$ and $\{m_3,m_4\}$ at $\bm{R}_j$ and $\Braket{ij}$ represents a summation for Ce-Ce vectors $\bm{R}_{ij}=\bm{R}_i-\bm{R}_j$ and $K_{m_1m_2m_3m_4}(\bm{q})$ is given by,
\begin{align}
K_{m_1m_2m_3m_4}(\bm{q})=\frac{1}{N}\sum_{\bm{k}ss\rq{}}
\frac{
v_{\bm{k}s}^{m_3m_1}
v_{\bm{k}+\bm{q}s\rq{}}^{m_2m_4}
}{(\mu-\varepsilon_{\Gamma_8}^{f})^2}
\frac{
f(\varepsilon_{\bm{k}+\bm{q}s\rq{}}^{c})-f(\varepsilon_{\bm{k}s}^{c})}{
\varepsilon_{\bm{k}s}^{c}-\varepsilon_{\bm{k}+\bm{q}s\rq{}}^{c}},
\label{eq:Kq}
\end{align}
where $f(x)$ is the Fermi distribution function $f(x)=1/(e^{(x-\mu)/T}+1)$ and $\mu$ is a chemical potential. 
Here $v_{\bm{k}s}^{mm\rq{}}$ is a $f$-$c$ mixing matrix between $m$ and $m\rq{}$ via the $c$ band state with $\bm{k},s$ given by, 
\begin{align}
v_{\bm{k}s}^{mm\rq{}}=\sum_{\ell\ell\rq{}}
V_{\bm{k}m\ell}^{*}
V_{\bm{k}m\rq{}\ell\rq{}}
u_{\bm{k}s\ell}^{c*}
u_{\bm{k}s\ell\rq{}}^{c},
\end{align}
%which is a key quantity including whole information about the $f$ state scattering between $\{m,m\rq{}\}$ through the $c$ state with $\bm{k},s$. 
which has all information about the $f$ state scattering between $\{m,m\rq{}\}$ through the $c$ state with $\bm{k},s$. 

The multipole interaction in $\bm{q}$-space %, which is important for the actual multipole ordering, 
is explicitly given by the follwing form, 
\begin{align}
&\overline{K}_{O_{\Gamma}}(\bm{q})
=\sum_{m_1m_2}\sum_{m_3m_4}O_{m_1m_2}^{\Gamma}O_{m_4m_3}^{\Gamma}
\left(K_{m_1m_2m_3m_4}(\bm{q})-K_{m_1m_2m_3m_4}^{\rm loc}\right)
\end{align}
where $O_{mm\rq{}}^{\Gamma}$ is the matrix element of the multipole operator %in $\Gamma_{8}$ basis 
and $K_{m_1m_2m_3m_4}^{\rm loc}=(1/N)\sum_{\bm{q}}K_{m_1m_2m_3m_4}(\bm{q})$. 
The mean-field multipole susceptibility $\chi_{O_{\Gamma}}^{}(\bm{q})$ is written by,
\begin{align}
\chi_{O_{\Gamma}}^{}(\bm{q})=\frac{
\chi_{O_{\Gamma}}^{0}(\bm{q})}{1-\chi_{O_{\Gamma}}^{0}(\bm{q})\overline{K}_{O_{\Gamma}}(\bm{q})},
\label{eq:chi_OMF}
\end{align}
which is enhanced towards the multipole ordering instability for the ordering moment $O_{\Gamma}$ and wavevector $\bm{q}=\bm{Q}$, 
and diverges at a critical point of the multipole ordering transition temperature $T=T_{O_{\Gamma}}^{\bm{Q}}$ 
where $\chi_{O_{\Gamma}}^{0}(\bm{q})\overline{K}_{O_{\Gamma}}(\bm{q})$ reaches unity. 
In the localized $f$ limit, the $\Gamma_{8}$ ground state is degenerate and the single-site susceptibility for all multipole moments exhibits the Curie law, $\chi_{O_{\Gamma}}^{0}(\bm{q})=1/T$, 
and then the transition temperature for a certain multipole ordering %for the multipole ordering moment $O_{\Gamma}$ and wavevector $\bm{Q}$ 
is determined by the condition $T_{O_{\Gamma}}^{\bm{Q}}=\overline{K}_{O_{\Gamma}}^{\rm max}(\bm{Q})$. 
Therefore 
%The $\bm{q}$-dependence of $\chi_{O_{\Gamma}}^{}(\bm{q})$ is weak as shown in Sec. \ref{sec3}, 
the sign and maximum value of $\overline{K}_{O_{\Gamma}}^{}(\bm{q})$ plays an central role for the multipole ordering. 
%determines the multipole ordering moment and wavevector for any given $T$. 
Hereafter we set $\mu-\varepsilon_{\Gamma_8}^{f}=2$ eV, % in order to compare the relistic coupling value, 
%since this factor is independent of $\bm{q}$ and $m$, and hence does no affect the ordering type and wavevector.
%,whose effect is discussed in Sec. \ref{sec5}. 
and $\mu$ is determined so as to keep $n_{tot}=n^{c}=21$ and $T$ is set to $T=0.005$ eV throughout the calculation. 

%Fig.2%%%%%%%%%%%%%%%%%%%%%%%%%%%%%%%%%%%%%%%%%%%%%%%%%%%%%%%%%%%%%%%%%%%%%%%%%%%%%%%%%%%%%%%%
\begin{figure*}[t]
\centering
\includegraphics[width=10.0cm]{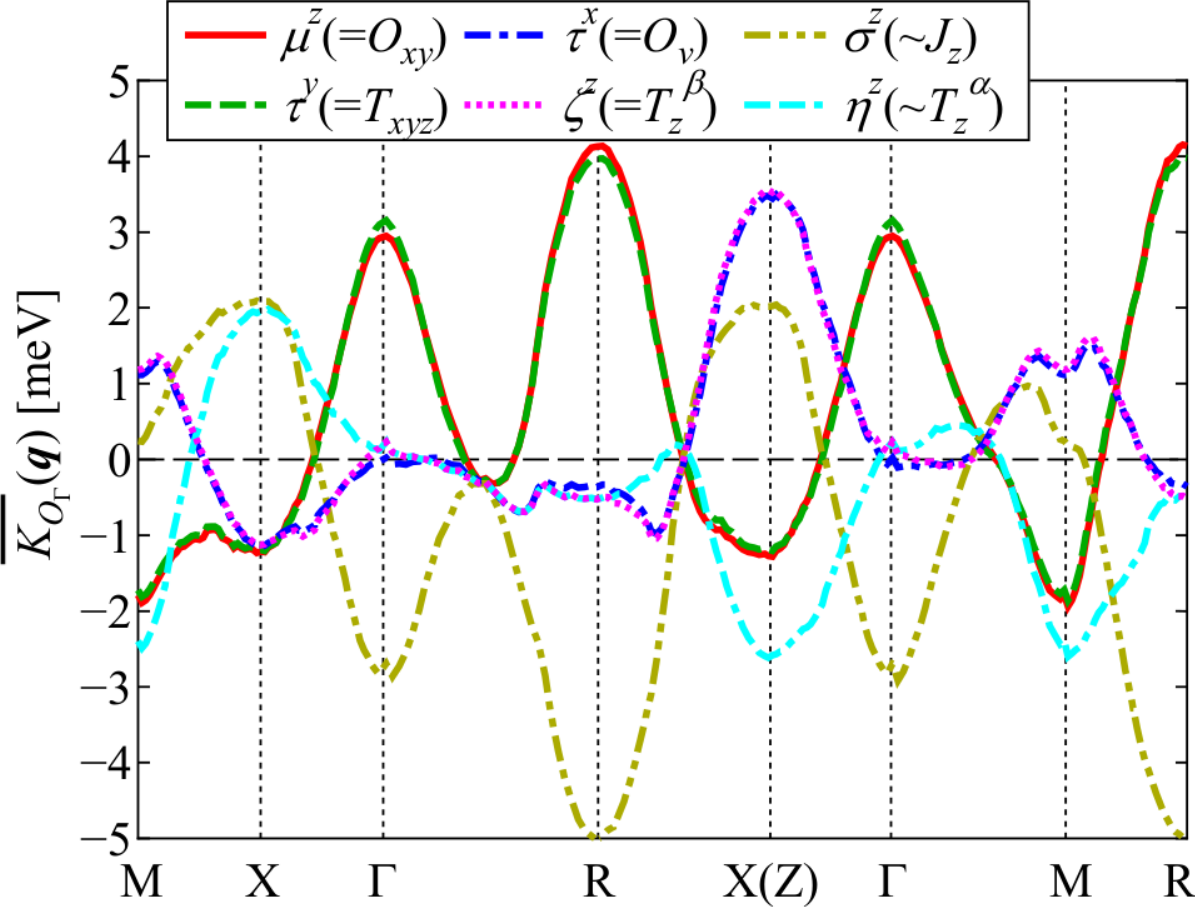}
%\vspace{-0.3cm}
\caption{(Color online) 
The $\bm{q}$-dependence of the RKKY multipole coupling $\overline{K}_{O_{\Gamma}}^{}(\bm{q})$ 
for the multipole moments of the typical irreducible representations. 
The individual terms for $\Gamma_{2u}$ octupole $T_{xyz}$, $\Gamma_{3g}$ quadrupoles $O_{u},O_{v}$, 
$\Gamma_{4u}$ multipoles $(\sigma^{x},\sigma^{y},\sigma^{z})$ and $(\eta^{x},\eta^{y},\eta^{z})$ and, 
$\Gamma_{5g}$ quadrupoles $(O_{yz},O_{zx},O_{xy})$ and $\Gamma_{5u}$ octupoles $(\zeta^{x},\zeta^{y},\zeta^{z})$. 
%$\bm{Q}_{1}=(\tfrac{\pi}{2},\tfrac{\pi}{2},0)$ and $\bm{Q}_{2}=(\tfrac{\pi}{2},\tfrac{\pi}{2},\pi)$ are the AFM ordering vectors of phase III. 
}
\label{Fig2}
%\vspace{+0.2cm}
\end{figure*}
%%%%%%%%%%%%%%%%%%%%%%%%%%%%%%%%%%%%%%%%%%%%%%%%%%%%%%%%%%%%%%%%%%%%%%%%%%%%%%%%%%%%%%%%%%%%%%%

\section{Results}
%\subsection{$\bm{q}$-dependence of RKKY coupling $\overline{K}_{O_{\Gamma}}^{}(\bm{q})$}
The RKKY couplings $\overline{K}_{O_{\Gamma}}^{}(\bm{q})$ for several multipole moments as a function of the wavevector $\bm{q}$ along the high symmetry line in the BZ 
are plotted as shown in Fig. \ref{Fig2}, 
where the positive (negative) coupling for a certain multipole with ($O_{\Gamma}$, $\bm{q}$) 
%gives the transition temperature towards the corresponding multipole instability. 
enhances (suppresses) the corresponding multipole fluctuation and its positive maximum value gives a leading multipole ordering mode. 
%The obtained results for the leading multipole ordering modes upto the 10th largest coupling 
%together with type of the multipole moments, ordering wavevectors and maximum values of couplings 
All leading modes are summarized in Table II of Ref. \cite{YH2019}. 

The couplings of the $\Gamma_{5g}$ quadrupole $O_{xy}$ and $\Gamma_{2u}$ ocutupole $T_{xyz}$ 
for $\bm{q}=(\pi,\pi,\pi)$ become largest among all moments and $\bm{q}$, 
which corresponds to the AFQ ordering of CeB$_6$ as phase II. 
%In addition, $\Gamma_{2u}$ ocutupole $T_{xyz}$ coupling is quite large and comparable to the $\Gamma_{5g}$ quadrupoles with the same wavevector as shown in Fig. \ref{Fig2} but slightly small. 
From the analysis of the real space couplings of $O_{xy}$ and $T_{xyz}$, 
we have found that the main origin of this mode comes from the fact that 
the couplings with the 1st and 2nd neighbor Ce-Ce vectors 
%1st and 2nd neighbor couplings between Ce-Ce vectors 
exhibit an anti-ferro (AF) and ferro (F) interaction respectively, 
which indicates the realization of the RKKY oscillation as shown in Fig. 7 in Ref. \cite{YH2019}
but the absolute value of the 2nd neighbor coupling is almost same or slightly larger than that of the 1st neighbor coupling. 
The 2nd neighbor F couplings of $O_{xy}$ and $T_{xyz}$ also increase the uniform mode with a substantial peak for $\bm{q}=(0,0,0)$ as shown in Fig. \ref{Fig2} corresponding to the elastic softening of $C_{44}$\cite{Luthi1984}. 

The difference of the couplings between $O_{xy}$ and $T_{xyz}$ has often been discussed in the early studies\cite{Shiba1999,Hanzawa2000}, 
where the two couplings must have the same value within the 1st neighbor coupling due to the point group symmetry. 
In the present calculation, %we have obtained 
the same coupling value of the 1st neighbor $O_{xy}$ and $T_{xyz}$ has been obtained, 
%and thus the early consideration is satisfied. 
while in the 2nd neighbor couplings, a slightly but finite difference between $O_{xy}$ and $T_{xyz}$ has been observed 
for the Ce-Ce vectors $\bm{R}=a(011),a(101)$ where $a$ is the lattice constant, 
which enhances the peak of $O_{xy}$ over that of $T_{xyz}$ at $\bm{q}=(\pi,\pi,\pi)$. 
The anisotropy of the present interaction Hamiltonian 
including the origin of the above correction 
together with the so-called \lq{}bond density\rq{}\cite{Shiba1999} 
will be presented  in the subsequent paper\cite{HY2019}. 
%which seems to be the same value from the previous discussions\cite{Shiba1999,Hanzawa2000} 
%where $O_{xy}$ and $T_{xyz}$ have almost same matrix elements and yield the similar fluctuations in phase I. 
%Furthermore the quadupoles $(O_{yz},O_{zx},O_{xy})$ and octupole $T_{xyz}$ take a substantial peak for $\bm{q}=(0,0,0)$, 
%which also corresponds to the elastic softening of $C_{44}$\cite{Luthi1984,Goto1985,Nakamura1994,Nakamura1995}. 
%Furthermore the quadupoles $(O_{yz},O_{zx},O_{xy})$ and octupole $T_{xyz}$ couplings %of the uniform mode with $\bm{q}=(0,0,0)$ 

The next largest coupling is the $\Gamma_{5u}$ octupole $\zeta^{z}$ at $\bm{q}=(0,0,\pi)$ [X(Z) point] 
which is degenerate for $\zeta^{x}~[\zeta^{y}]$ octupole at $\bm{q}=(\pi,0,0)~[(0,\pi,0)]$. 
In addition to this, the $\Gamma_{3g}$ quadrupole $O_{v}=O_{x^2-y^2}$ coupling is quite large for $\bm{q}=(0,0,\pi)$ and becomes similar value of the octupole coupling $\zeta^{z}$, 
which is also degenerate for the rotated moments to the each principle-axis $O_{y^2-z^2}$ and $O_{z^2-x^2}$
for $\bm{q}=(\pi,0,0)$ and $\bm{q}=(0,\pi,0)$ respectively.
In real space, the coupling of $\zeta^{z}$ and $O_{v}$ are highly anisotropic 
%the coupling with the intersite vector parallel to the moment direction
%the coupling with the intersite vector perpendicular to the moment direction
with the AF couplings for the Ce-Ce vector $\bm{R}=a(001)$ and the F couplings for $\bm{R}=a(100),a(010)$, 
which induces the enhancement of the $\bm{q}=(0,0,\pi)$ mode. 
In these situation, the triple-$\bm{Q}$ mode of $(\zeta^{x},\zeta^{y},\zeta^{z})$ and $(O_{y^2-z^2},O_{z^2-x^2},O_{x^2-y^2})$
with $\bm{Q}=(\pi,0,0),(0,\pi,0),(0,0,\pi)$ may become possible for the phase IV in Ce$_x$La$_{1-x}$B$_6$ with $x<0.8$, 
which is different from the $\bm{q}=(\pi,\pi,\pi)$ AFO ordering of $(\zeta^{x}+\zeta^{y}+\zeta^{z})/\sqrt{3}$ 
%for the phase IV observed in the La-doping system Ce$_x$La$_{1-x}$B$_6$ with $x<0.8$
\cite{Kubo-Kuramoto2004}. 
%The role of the $\Gamma_{5u}$ octupoles $(\zeta^{x},\zeta^{y},\zeta^{z})$ is also discussed for 
%the phase IV observed in the La-doping system Ce$_x$La$_{1-x}$B$_6$ with $x<0.8$\cite{Cameron2016}, 
%where the $\bm{q}=(\pi,\pi,\pi)$ antiferro-octupolar (AFO) ordering of $(\zeta^{x}+\zeta^{y}+\zeta^{z})/\sqrt{3}$ is considered to be a possible mode. 
%In contrast, the present result indicates 
%the $\Gamma_{5u}$ AFO with the domained structure of $\zeta^{x}$, $\zeta^{y}$ and $\zeta^{z}$ for $\bm{q}=(\pi,0,0)$, $(0,\pi,0)$ and $(0,0,\pi)$, respectively. 
The present development of the $\zeta^{z}$ and $O_{v}$ X-point mode 
appears to be related to the recent inelastic neutron scattering experiments in Ce$_{x}$La$_{1-x}$B$_6$\cite{Nikitin2018}, 
where the intensity $\bm{q}=(\pi,0,0)$ is enhanced and becomes dominant mode for $x<0.8$.

As for the phase III of the AFM order, the $\Gamma_{4u}$ magnetic multipole couplings of $\sigma^{z}$ and $\eta^{z}$ 
shall be dominant when the system enters into phase II.
They does not become so large in the present paramagnetic system (phase I)
and their maximum values are less than half of the 1st leading peak value of $\Gamma_{5g}$-$(\pi,\pi,\pi)$. 
The realistic description of the successive transition from phase I (paramagnetic) to phase II (AFQ) and from phase II (AFQ) to phase III (AFM) is an important future problem.

\section{Summary}
In summary, we have performed a direct calculation of the RKKY interactions based on the 74-orbital effective Wannier model derived from the bandstructure calculation of CeB$_6$. 
%When we drop the $f$-$f$ hopping and $f$-$c$ mixing of the Wannier model, the $c$ band dispersion is almost the same as the GGA+$U$-band of LaB$_6$ with a single ellipsoidal FS centered at X point. This is in good agreement with the recent ARPES\cite{Neupane2015,Ramankuttya2016,Koitzsch2016} and early dHvA experiments\cite{Onuki1989}. 
We obtain the RKKY couplings for the active multipole moments in $\Gamma_8$ subspace explicitly
as functions of wavevector $\bm{q}$. %and inter-cell vector $\bm{R}_{ij}$. 
%, where we derive a useful expression in order to treat all 60 $c$-orbital contributions. 
The couplings of the $\Gamma_{5g}$ quadrupole $O_{xy}$ together with the $\Gamma_{2u}$ octupole $T_{xyz}$ are highly enhanced for $\bm{q}=(\pi,\pi,\pi)$ and $\bm{q}=(0,0,0)$ %as the 5th and 6th leading modes, 
where the former explains the AFQ ordering of the phase II and the latter corresponds to the elastic softening of $C_{44}$. 
%The 3rd (4th) leading mode is the $\Gamma_{5u}$-AFO ($\Gamma_{3g}$-AFQ) of $\zeta^{x}$, $\zeta^{y}$ and $\zeta^{z}$ ($O_{y^2-z^2}$, $O_{z^2-x^2}$ and $O_{x^2-y^2}$) with the corresponding wavevectors for $\bm{q}=(\pi,0,0)$, $(0,\pi,0)$ and $(0,0,\pi)$ respectively, which are almost degenerate each other and differ from the discussed AFO-mode with $(\zeta^{x}+\zeta^{y}+\zeta^{z})/\sqrt{3}$ at $\bm{q}=(\pi,\pi,\pi)$\cite{Kubo-Kuramoto2003,Kubo-Kuramoto2004,Mannix2005,Kurasawa2007,Matsumura2014,Inami2014,Sera2018}. 
%All the obtained RKKY couplings have long-range and oscillating behavior as a function of $\bm{R}_{ij}$, where the $\Gamma_{5g}$ quadrupole $(O_{yz},O_{zx},O_{xy})$ and $\Gamma_{2u}$ octupole $T_{xyz}$ couplings indicate the sign-reversing for each neighboring site and have a positive largest value at the second neighbor which cooperatively enhances the AFQ with $\bm{q}=(\pi,\pi,\pi)$, while for the second leading $\Gamma_{5u}$ AFO mode, the anisotropic first neighbor couplings are significant. This induces the leading mode shift with increasing the La-substitution rate $x$ in Ce$_x$La$_{1-x}$B$_6$ from the $(\pi,\pi,\pi)$-AFQ with $O_{xy}$ (phase II) to the $(\pi,0,0)$-AFO with $\zeta^{x}$ (phase IV) which may be also consistent with the $(\pi,0,0)$ peak in the INS data\cite{Nikitin2018}. 
The present approach enables us to access the possible multipole ordering moments and wavevectors without any assumption and to provide a good insight for searching the multipole ordering in connect with the inherent feature and the concrete situation of actual compounds. 
% such as the changes of FSs, carrier densities, lattice constants and internal coordinates of atoms. 

%\section*{References}
%\bibliography{SCES2019-proc-Ce-RKKY}

\end{document}